\documentclass[twocolumn,amsmath,amssymb,aps,prb,showpacs,showkeys,superscriptaddress,nobibnotes,footinbib]{revtex4-1}

\usepackage{graphicx}
\usepackage{dcolumn}
\usepackage{bm}
\usepackage{braket,amsmath,amssymb,bm}
\usepackage{color}
\usepackage{siunitx}
\usepackage{float}
\usepackage{hyperref}
\usepackage{lineno}
\usepackage{xcolor}

\makeatletter

\begin{document}

\title{Real-time nanodiamond thermometry probing \textit{in-vivo} thermogenic responses}

\author{Masazumi Fujiwara}
 \thanks{masazumi@osaka-cu.ac.jp}
  \affiliation{Department of Chemistry, Osaka City University, Sumiyoshi-ku, Osaka, 558-8585, Japan}

\author{Simo Sun}
 \affiliation{Food and Human Health Sciences, Graduate School of Human Life Science, Osaka City University, Sumiyoshi-ku, Osaka, 558-8585, Japan}
 
\author{Alexander Dohms}
 \affiliation{Institut f{\" u}r Physik und IRIS Adlershof, Humboldt Universit{\" a}t zu Berlin, Newtonstrasse 15, 12489 Berlin, Germany}

\author{Yushi Nishimura}
  \affiliation{Department of Chemistry, Osaka City University, Sumiyoshi-ku, Osaka, 558-8585, Japan}
 
\author{Ken Suto}
  \affiliation{Department of Chemistry, Osaka City University, Sumiyoshi-ku, Osaka, 558-8585, Japan}

\author{Yuka Takezawa}
 \affiliation{Food and Human Health Sciences, Graduate School of Human Life Science, Osaka City University, Sumiyoshi-ku, Osaka, 558-8585, Japan}

\author{Keisuke Oshimi}
  \affiliation{Department of Chemistry, Osaka City University, Sumiyoshi-ku, Osaka, 558-8585, Japan}

\author{Li Zhao}
 \affiliation{State Key Laboratory of Radiation Medicine and Protection, School for Radiological and Interdisciplinary Sciences (RAD-X) and Collaborative Innovation Center of Radiation Medicine of Jiangsu Higher Education Institutions, Soochow University, Suzhou 215123, P. R. China}

\author{Nikola Sadzak}
 \affiliation{Institut f{\" u}r Physik und IRIS Adlershof, Humboldt Universit{\" a}t zu Berlin, Newtonstrasse 15, 12489 Berlin, Germany}
 
\author{Yumi Umehara}
  \affiliation{Department of Chemistry, Osaka City University, Sumiyoshi-ku, Osaka, 558-8585, Japan}

\author{Yoshio Teki}
  \affiliation{Department of Chemistry, Osaka City University, Sumiyoshi-ku, Osaka, 558-8585, Japan}
 
\author{Naoki Komatsu}
 \affiliation{Graduate School of Human and Environmental Studies, Kyoto University, Sakyo-ku, Kyoto, 606-8501, Japan}

\author{Oliver Benson}
 \affiliation{Institut f{\" u}r Physik und IRIS Adlershof, Humboldt Universit{\" a}t zu Berlin, Newtonstrasse 15, 12489 Berlin, Germany}

\author{Yutaka Shikano}
\thanks{yutaka.shikano@keio.jp}
 \affiliation{Quantum Computing Center, Keio University, 3-14-1 Hiyoshi, Kohoku, Yokohama, 223-8522, Japan}
 \affiliation{Institute for Quantum Studies, Chapman University, 1 University Dr., Orange, CA 92866, USA}

 \author{Eriko Kage-Nakadai}
\thanks{nakadai@life.osaka-cu.ac.jp
\\ M.F., S.S., A.D. contributed equally to this work.}
 \affiliation{Food and Human Health Sciences, Graduate School of Human Life Science, Osaka City University, Sumiyoshi-ku, Osaka, 558-8585, Japan}

\begin{abstract}
\textbf{
Real-time temperature monitoring inside living organisms provides a direct measure of their biological activities, such as homeostatic thermoregulation and energy metabolism.
However, it is challenging to reduce the size of bio-compatible thermometers down to submicrometers despite their potential applications for the thermal imaging of subtissue structures with single-cell resolution.
Light-emitting nanothermometers that remotely sense temperature via optical signals exhibit considerable potential in such \textit{in-vivo} high-spatial-resolution thermometry.
Here, using quantum nanothermometers based on optically accessible electron spins in nanodiamonds (NDs), we demonstrate \textit{in-vivo} real-time temperature monitoring inside \textit{Caenorhabditis elegans} (\textit{C. elegans}) worms.
We developed a thermometry system that can measure the temperatures of movable NDs inside live adult worms with a precision of $\pm$ 0.22 $\si{\degreeCelsius}$.
Using this system, we determined the increase in temperature based on the thermogenic responses of the worms during the chemical stimuli of mitochondrial uncouplers.
Our technique demonstrates sub-micrometer localization of real-time temperature information in living animals and direct identification of their pharmacological thermogenesis.
The results obtained facilitate the development of a method to probe subcellular temperature variation inside living organisms and may allow for quantification of their biological activities based on their energy expenditures.
}
\end{abstract}

% \pacs{07.20Dt, 61.72.Ji, 42.30.}
% %07.20.Dt Thermometers
% %61.72.Ji Point defects (vacancies, interstitials, color centers, etc.) and defect clusters
% %42.30.: Imaging and optical processing
% \keywords{nanodiamond, thermometry, in-vivo, quantum}
\maketitle

\section*{Introduction}
\begin{figure*}[t!]
 \centering
 \includegraphics[scale=1.0]{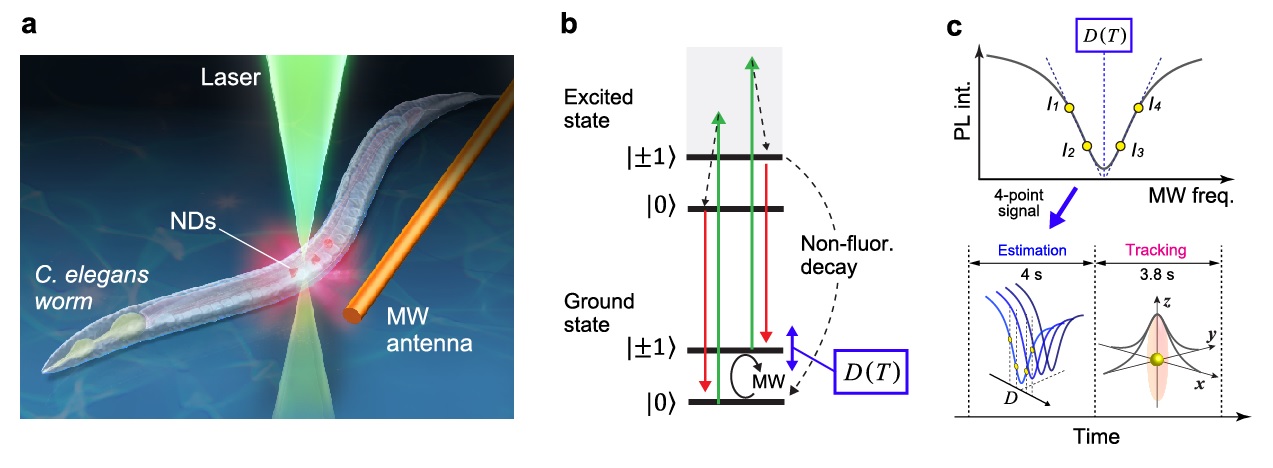}
 \caption{\textbf{Real-time nanodiamond thermometry of \textit{C. elegans} worms.} 
 (a) ND quantum thermometers probing inside the worms. NDs are incorporated in the worms. ODMR of NV centers can be observed by applying a green laser and microwave excitation. (b) A simplified energy diagram of the excited and ground states of NV centers with the associated electron spin states. Green and red arrows indicate the laser excitation and fluorescence, respectively. Microwave (MW) excites the spin state $\ket{0} \rightarrow \ket{\pm 1}$ in the ground state, which are separated by temperature-dependent zero-field splitting ($D(T)$).
 The following optical excitation initializes the spin state to the ground-state $\ket{0}$ through non-fluorescent decay (dashed arrow) from the excited-state $\ket{\pm 1}$.
 (c) An experimental scheme to estimate $D(T)$ as temperature with fast particle tracking. $D(T)$ is estimated from the spin-dependent fluorescence intensities $I_1$ to $I_4$ at the four frequency points on the ODMR spectrum.
}
 \label{fig1}
\end{figure*}

The temperature inside living organisms is a direct measure of their biological activities.
A poikilotherm is temperature-dependent organism, and even a homeotherm shows internal temperature variations under normal physiological conditions, as can be seen in homeostatic thermoregulation~\cite{akin2011homeostatic} and energy metabolism~\cite{10.3389/fphys.2017.00520}.
Their sub-micrometer-scale temperature information should provide rich information on cellular and molecular activities that has potential applications for the thermal imaging of brain subtissue structures~\cite{10.3389/fncom.2016.00082}, thermal visualization of intratumor heterogeneity~\cite{gorbach2004intraoperative,marusyk2012intra}, and thermogenic mapping of adipocytes~\cite{van2009cold}.
It is, however, challenging to reduce the size of bio-compatible thermometers down to submicrometers.
Conventional electric thermometers do not have submicrometer-scale resolution, and near-infrared thermography generally helps determine the surface temperature of biological specimens~\cite{LAHIRI2012221}.
Light-emitting nanothermometers, such as thermo-responsive molecular probes
~\cite{kriszt2017optical,kiyonaka2013genetically,donner2012mapping} and nanoparticles~\cite{yang2011quantum,del2018vivo}, may resolve these technical limitations. 
They were first developed for \textit{in-vitro} cultured cells~\cite{kriszt2017optical,kiyonaka2013genetically,donner2012mapping,yang2011quantum} and recently \textit{in-vivo} model animals~\cite{del2018vivo}.
The technical challenges of their \textit{in-vivo} applications are improving their long-term robustness, enabling them to follow the relatively slow response of body temperatures for hours~\cite{10.3389/fphys.2017.00520,del2018vivo}, and ensuring their nontoxicity at the  high dosages necessary for  \textit{in-vivo} measurements. 

The ND quantum thermometers considered in the present study have emerged as a promising candidate~\cite{RevModPhys.89.035002} exhibiting ultimate robustness~\cite{Chipaux2018small,hui2019carbon}, ultralow toxicity~\cite{hui2019carbon,mohan2010vivo,simpson2017non}, and quantum-enhanced high sensitivity in living cells~\cite{kucsko2013nanometre,simpson2017non}.
The sensor reads temperature as a frequency shift of the optically detected magnetic resonance (ODMR) of nitrogen-vacancy (NV) defect centers, which originates from thermal lattice expansion. 
The NV sensory core is deeply embedded in the diamond lattice and immune to various biological environmental factors~\cite{zhang2018hybrid,fujiwara2019monitoring}.
Implementing this quantum sensor in organisms will enable monitoring their site-specific thermal activities in real-time.

\section*{Results}
Our \textit{in-vivo} thermometry system is based on a confocal fluorescence microscope equipped with a microwave irradiation setup, and it is optimized for measuring biological specimens (Fig.~\ref{fig1}a).
The ODMR of NV centers can be measured as a decrease in the laser-induced fluorescence intensity when spin-resonant microwave excitation is applied, because the spin excitation activates the nonfluorescent relaxation pathway from the excited state to the ground state (Fig.~\ref{fig1}b).
Further, this ODMR confocal microscope performs fast particle tracking and real-time high-precision temperature estimation from the ODMR shift of NV centers.
In particle tracking, the system measures the ND fluorescence intensity along the microscope $xyz$ axes and focuses on the respective fluorescence maximum every 4 s, during which temperature is estimated with a sampling time of 0.5--1.0 s based on a 4-point ODMR measurement protocol described previously~\cite{kucsko2013nanometre} (Fig.~\ref{fig1}c). 
This estimation implicitly assumes that photon counts registered at all the four selected frequencies are linearly scaled to the changes in the detected fluorescence intensity. 
However, each photon-count number exhibits marginal differences in the photo-responsivity, a low percentage (see Fig.~\ref{figS-calib}), which actually creates significant artefacts in the real-time frequency-shift estimate, particularly in the low-photon regime.
We therefore applied an error correction filter to cancel these effects.

\begin{figure}[h!]
 \centering
 \includegraphics[scale=1]{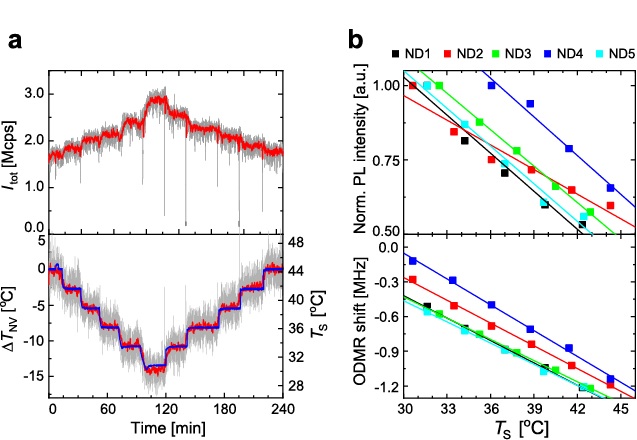} 
 \caption{\textbf{Basic characterization of ND fluorescence intensity and ODMR shift.}  
 (a) Time profiles of total photon counts $I_{\rm tot}$ (top), and  $\Delta T_{\rm NV}$ (bottom) over the stepwise temperature variation of $T_{\rm S}$.
 $\Delta T_{\rm NV}$ is calculated with $dD /dT = - 65.4 \ \si{\kHz} \cdot \si{\degreeCelsius}^{-1}$ experimentally determined in Fig.~\ref{fig2}b.
 Gray: $\Delta T_{\rm NV}$ of every 1 s. Red: moving average of 20 sampling points. Blue: $T_{\rm S}$. 
 (b) Temperature dependence ($T_{\rm S}$) of normalized PL intensity (top) and ODMR shift (bottom) of five NDs on coverslips.
}
 \label{fig2}
\end{figure}

Figure~\ref{fig2}a shows the time profiles of the total photon counts ($I_{\rm tot}$) and temperature estimate of the NDs ($\Delta T_{\rm NV}$) when 
the temperature of the sample ($T_{\rm S}$) is varied from $44.3 \rightarrow 30.4 \rightarrow 44.3 \si{\degreeCelsius}$ in steps of $\sim$ 2.8 $\si{\degreeCelsius}$.
Our system now accurately provides $\Delta T_{\rm NV}$ corresponding to $T_{\rm S}$ 
(see Methods for details on how to vary and calibrate $T_{\rm S}$).
In addition, $\Delta T_{\rm NV}$ clearly demonstrates anti-correlation with $I_{\rm tot}$ as reported previously~\cite{plakhotnik2010luminescence}.
A statistical study on this type of temperature dependency determines the means and standard deviations for $I_{\rm tot}^{-1} \ dI_{\rm tot}/dT = -3.9 \pm 0.7 \ \% \cdot \si{\degreeCelsius}^{-1}$ and $dD/dT  =- 65.4 \pm 5.5 \ \si{\kHz} \cdot \si{\degreeCelsius}^{-1}$,
which slightly differ from the previously reported values~\cite{plakhotnik2010luminescence}.
The variation of $dD/dT$ results in a $\Delta T_{\rm NV}$ error of $\sim$ 8 \% of the measured $\Delta T_{\rm NV}$ (see Supplementary Information).
The temperature precision and accuracy  (r.m.s.) are $\pm$ 0.24 $\si{\degreeCelsius}$ and $<$ 0.6 $\si{\degreeCelsius}$, respectively, giving a sensitivity of 1.5 $\si{\degreeCelsius}/\sqrt{\si{Hz}}$ (see Methods).
Strikingly, the thermometry detects only a very small influence of the environmental temperature drift ($T_{\rm air}$) of $\pm$ 0.1 $\si{\degreeCelsius}$ (Fig.~\ref{figS-stab}).
We also verify the particle tracking effectivity as it rapidly tracks the positional shift of the focus during the step variation; 3 $\si{\degreeCelsius}$ causes a drastic $z$-positional shift of 6 $\si{\um}$ for 3--4 min ($78 \ \rm{nm} \cdot {\rm s^{-1}}$ at maximum, see Fig.~\ref{figS-precacc}c).

\begin{figure}[t!]
 \centering
 \includegraphics[scale=1.1]{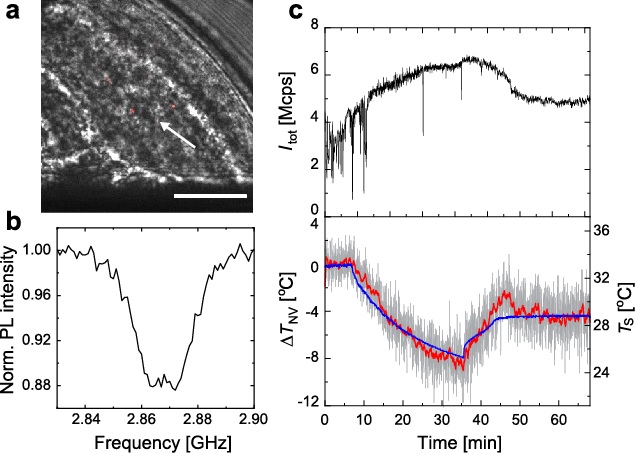}
 \caption{\textbf{\textit{In-vivo} temperature measurement in \textit{C. elegans} worms with environmental temperature changes.} (a) A red-gray merged photo of NDs in the worm. Red-scale: red fluorescence. Gray-scale: bright-field. The white arrow indicates the ND used for the temperature measurements. The black shadow seen in the bottom part of the image is the microwave linear antenna. Scale bar: 20 $\si{\um}$.
 (b) CW-ODMR spectrum of the ND.  
 (c) Time profiles of $I_{\rm tot}$ (top) and $\Delta T_{\rm NV}$ (bottom) during temperature change. 
 Gray: $\Delta T_{\rm NV}$ reading every 1 s, Red: moving average of 20 sampling points, Blue: $T_{\rm S}$. $\Delta T_{\rm NV}$ is calculated with $dD /dT = - 65.4 \ \si{\kHz} \cdot \si{\degreeCelsius}^{-1}$.
 }
 \label{fig4}
\end{figure}

Having established robust and accurate thermometry for real-time operation, we test the local temperature monitoring in live worms while applying a temperature shock in the context of a thermosensation study.~\cite{prahlad2008regulation}
%}
%
Figure~\ref{fig4}a shows a picture of the NDs inside worms that are anesthetized and placed near the microwave antennae.
These NDs are highly water-dispersible by the surface functionalization of polyglycerol (PG)~\cite{zhao2011chromatographic} and are introduced by microinjection into the gonads~\cite{mohan2010vivo} (see Methods).
Figure~\ref{fig4}b is a CW-ODMR spectrum of a single ND, which is denoted by the arrow in Fig.~\ref{fig4}a. 
Figure~\ref{fig4}c shows the time profiles of $I_{\rm tot}$ and $\Delta T_{\rm NV}$ over a period of one hour during a temperature change of $T_{\rm S}$. 
We begin measurements for $T_{\rm obj}$ at 33.2 $\si{\degreeCelsius}$ and decrease it to 25.3 $\si{\degreeCelsius}$ at 6 min.
It is subsequently set to 28.6 $\si{\degreeCelsius}$ at 35.2 min. 
$\Delta T_{\rm NV}$ accurately gives the temperature change between the two stationary states of 33.2 and 28.6 $\si{\degreeCelsius}$. 
The \textit{in-vivo} precision and accuracy values are $\pm$ 0.22 $\si{\degreeCelsius}$ (gradually varies to 0.31 $\si{\degreeCelsius}$) and $<$ 0.6 $\si{\degreeCelsius}$, respectively (Fig.~\ref{figS-invivo-prec}), with a sensitivity of 1.4 $\si{\degreeCelsius}/\sqrt{\si{Hz}}$. 
Between these two stationary states, the temperature dynamics inside worms are reflected because $\Delta T_{\rm NV}$ always lags behind $T_{\rm S}$ owing to the finite heat capacity of the microscope objective and worm surroundings, including the agar pads and buffer. 
$I_{\rm tot}$ also show temperature-induced gradual changes in fluorescence intensities. 
Particle tracking works effectively during measurement, as can be seen in Fig.~\ref{fig4}c; during $t =$ 0--15 min, the photon counts exhibit frequent spikes originating from ND positional fluctuations of approximately 400 nm for several seconds (Fig.~\ref{figS-invivo-prec} and Supplementary Information).
The present demonstration for accurate internal temperature measurement in worms can be directly used to quantify the heat (cold) shock of thermosensory neurons in combination with calcium imaging and optogenetics~\cite{doi:10.1111/boc.201200069}.

\begin{figure*}[t!]
 \centering
 \includegraphics[scale=1.1]{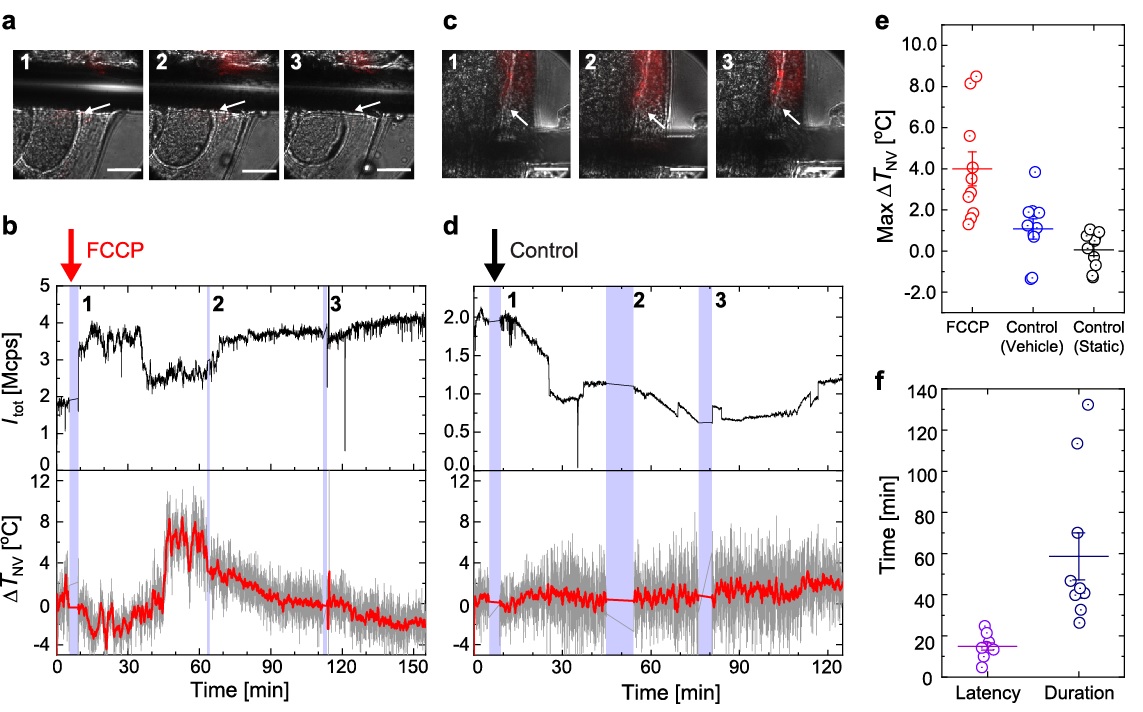}
 \caption{\textbf{Temperature rise inside \textit{C. elegans} worms by chemical stimulation.} (a, c) Merged photos of NDs during FCCP stimulation (60 $\mu$M) and vehicle-control experiments. Red-scale: red fluorescence. Gray-scale: bright-field. The numbers indicate the timestamps of pictures captured during measurement indicated in Figs.~\ref{fig5}b and d. Scale bar: 20 $\si{\um}$. (b, d) Time profiles of $I_{\rm tot}$ (top) and $\Delta T_{\rm NV}$ (bottom) during FCCP stimulation and vehicle-control experiments. The blue shaded regions represent periods when no temperature measurement is performed. The photographs in a, c are obtained during these periods. 
 $\Delta T_{\rm NV}$ is calculated with $dD /dT = - 65.4 \ \si{\kHz} \cdot \si{\degreeCelsius}^{-1}$ for both types of the experiments.
 (e) Statistical plots of the maximum $\Delta T_{\rm NV}$ for FCCP stimulation (red), vehicle-control (blue), and static-control (black, no solution added). $n$ = 10 for all data. The mean values with standard error are $4.0 \pm 0.8$, $1.1 \pm 0.5$, and $0.1 \pm 0.3 \si{\degreeCelsius}$ for the FCCP, vehicle-control, and static-control experiments, respectively. All measurements were performed at room temperature with a long-term fluctuation of $\pm 0.5 \si{\degreeCelsius}$ (see Fig.~\ref{figS-Tscalib}).
 (f) Latency and duration of the temperature-increase responses of $\Delta T_{\rm NV}$ for the FCCP stimulus, whose means with standard errors are $13.6 \pm 2.1$ and $54.9 \pm 10.9$ min, respectively. 
}
 \label{fig5}
\end{figure*}

As we have demonstrated the capability of quantifying the local temperature inside live worms, we use the ND thermometry for \textit{in-vivo} thermogenic studies.
In particular, we monitor the internal temperatures of worms under pharmacological treatment to induce non-shivering thermogenesis by using a mitochondrial uncoupler, i.e., carbonyl cyanide-\textit{p}-trifluoromethoxyphenylhydrazone (FCCP)~\cite{kriszt2017optical,krenger2018dynamic}.
Figures~\ref{fig5}a, b show sequence of microscopic images of NDs in worms when the worms are stimulated by FCCP and the time profile of $\Delta T_{\rm NV}$ of the arrow-indicated ND, respectively.
In this temperature response curve, at $t \sim 7$ min, the FCCP solution was added to the culture medium.
At $t \sim$ 32 min, $\Delta T_{\rm NV}$ starts to gradually increase; at $t \sim 48$ min, an additional increase is observed as the total change reaches 4--7 $\si{\degreeCelsius}$. 
This temperature rise lasts about 80 min until $t \sim$ 120 min. 
The observed anti-correlation between $I_{\rm tot}$ and $\Delta T_{\rm NV}$ also supports the temperature increase ($t$ = 40--70 min). 
During the stimulation, the NDs slowly move several micrometers over an hour, which corroborates the results of separate experiments in which the NDs were continuously observed under a commercial microscope (Fig.~\ref{figS-tracking}).
As a control experiment for the stimulation, we test a vehicle solution (Figs.~\ref{fig5}c, d).
At $t \sim 6$ min, the vehicle solution is added; however, $\Delta T_{\rm NV}$ exhibits a flat response, and does not exhibit any noticeable increase. 

To further confirm the temperature increase by FCCP, we measure the number of ND-labeled worms for the FCCP-treatment and control experiments, as shown in Fig.~\ref{fig5}e.
We observe the clear tendency for internal temperature increase when stimulated by FCCP, compared with the vehicle control experiments in which only the vehicle solution is added.
Another control experiment (static control) in which no buffer solution is added and $\Delta T_{\rm NV}$ is just monitored statically indicates that droplet addition causes the fluctuation of $\Delta T_{\rm NV}$ at a certain level via either temperature change or ODMR-shift artefacts; however, it is insufficient to create the observed ODMR shift with FCCP addition, further confirming the temperature increase by FCCP (see also Fig.~\ref{figS-correlation} for the three distinct types of clusters in the correlation plots of maximum $\Delta T_{\rm NV}$ and integrated area of the response curves.)
Figure~\ref{fig5}f summarizes the latency and duration of the observed temperature responses.
The wider distribution of duration when compared with that of latency may be caused by insufficient controllability of the FCCP concentration or dehydration-induced viability.
The quantitative interpretation of these response curves should consider several factors including data extraction from the temperature time trace, particle inhomogeneity of $dD/dT$ of NDs and lack of information on thermogenesis in \textit{C. elegans}, which are discussed in the Supplementary Information.

\section*{Discussion}
This study demonstrates the possibility of real-time temperature measurements at a submicrometer-scale resolution in living organisms.
Direct extensions of the present results are the recent ND-labeled animal models, including zebrafish and fruit fly models~\cite{hui2019carbon}.
Monitoring temperature during their embryogenesis and metamorphosis can provide information about ultradian rhythms or thermoregulation of the developmental processes~\cite{lucchetta2005dynamics,ronny2019effect}.
Indeed, in embryos of \textit{C. elegans}, the microscopic mechanism of the cell division cycle has been addressed by measuring the local temperature using ND quantum thermometry~\cite{misha}.
The maximum tracking speed of our system is 80 ${\rm nm} \cdot {\rm s^{-1}}$, which covers the crawling speeds of various cells ~\cite{MAIURI2012R673}.
For example, by combining the proposed system with \textit{intravital} microscopy~\cite{condeelis2003intravital}, it would be possible to probe site-specific temperatures of tumors, organs, and model organoids when ND-labeled cells migrate into them; such a process may be used to analyze the dynamics of cancer metastasis or stem-cell engraftment.

We also anticipate further scale-up to \textit{in-vivo} mammalian thermogenesis.
ND thermometers may probe micrometer-scale local temperatures inside or on the surface of organs in anesthetized rodents. 
Spatial mapping of thermal heterogeneity in brain tissues may enable an understanding of neurovascular coupling through metabolic heat generation~\cite{10.3389/fncom.2016.00082}, which can provide a more direct measure of neural activity than local oxygen consumption~\cite{10.3389/fnins.2018.00550} that can distinguish between anaerobic and aerobic metabolic processes~\cite{doi:10.1113/jphysiol.2003.048082}.
Other applications such as detecting the temperatures of individual organs during hyperthermia or hypothermia~\cite{10.3389/fphys.2017.00520} and quantifying the metabolic rates of adipocyte phenotypes in subcutaneous tissues~\cite{van2009cold,ikeda2017ucp1} are also envisaged. 

In conclusion, we have developed an ND quantum thermometry system that can measure the temperatures of movable NDs inside live adult worms of \textit{C. elegans} with a precision of $\pm$0.22 $\si{\degreeCelsius}$.
This \textit{in-vivo} thermometry accurately measured the temperature dynamics inside the worms during environmental temperature changes.
By using this system, we have determined the temperature increase caused by the worm's thermogenesis under the treatment of mitochondrial uncoupler stimuli.
The results highlight the potential to probe subcellular temperature variations inside living organisms and may allow for the study of the submicrometer thermal effects to biological processes.

\clearpage
\section*{Methods}
\subsection*{Real-time thermometry based on ODMR of NV centers}
The thermometry system is based on a confocal fluorescence microscope equipped with microwave control for ODMR measurements (Fig.~\ref{figS-exsetup}a).
A continuous-wave 532-nm laser was used for the excitation of NDs with an intensity in the range of 1--10 kW$\cdot {\rm cm ^{-2}}$ to adjust $I_{\rm tot}$ to be around 2 Mcps. 
An oil-immersion microscope objective with a numerical aperture of 1.4 was used for both the excitation and the fluorescence collection. 
The NV fluorescence was filtered by a dichroic beam splitter (Semrock, FF560-FDi01) and long pass filter (Semrock, BLP01-635R-25) to remove the residual green laser scattering and \textit{in-vivo} background fluorescence.
The fluorescence was then coupled to an optical fiber (Thorlabs, 1550HP) to be detected by an avalanche photodiode (Excelitas, SPCM AQRH-14). 
Samples were placed in the incubation chamber and the chamber was mounted on a piezo stage that enabled raster scanning and particle tracking. 
The APD output was fed to a data-acquisition board system (National Instruments, USB-6343 BNC, USB-6229 BNC), where four of the six equipped counters were used.

To implement both the CW- and 4-point- ODMR measurements, a stand-alone microwave source (Rohde \& Schwarz, SMB100A) and three USB-powered microwave sources (Texio, USG-LF44) were connected to an SP6T switch with a switching time of 250 ns (General Microwave, F9160).
The microwave was then amplified (Mini-circuit, ZHL-16W-43+) and fed to a microwave linear antenna placed on a coverslip (25-\si{\um}-thin copper wire) that was sealed with a home-made cell-culture dish possessing a hole in the center (Fig.~\ref{figS-exsetup}b, c).  
The typical microwave excitation power was estimated to be 33.7 mW (15.3 dBm) by considering the input power (30 dBm) and irradiation area calculated by the finite-element method (COMSOL) (Figs.~\ref{figS-exsetup}d--e). 
In the CW-ODMR measurements, APD detection was gated for microwave irradiation ON and OFF using the SP6T switch and a bit pattern generator (Spincore, PBESR-PRO-300), where the gate width was 200 \si{\us} for both gates, followed by a laser shut-off time of 100 $\si{\us}$, resulting in $I_{\rm PL}^{\rm ON}$ and $I_{\rm PL}^{\rm OFF}$ with a repetition rate of 2 kHz (Fig.~\ref{figS-exsetup}f).
Note that an external magnetic field was not applied in this study. 
In the 4-point-ODMR measurements, APD detection was gated for the corresponding microwave frequencies ($\omega_1$ to $\omega_4$), where the gate width for all four gates was 100 \si{\us}, each followed by an interval of 5 $\si{\us}$ (Fig.~\ref{figS-exsetup}g). 
This gated photon counting of the four counters was performed approximately 2380 times for a second.

The temperature dependence data of Fig.~\ref{fig2} and Fig.~\ref{fig4}c were measured for the spin-coated NDs (Ad\'{a}mas Nanotechnologies, NDNV100nmHi10ml, 500 NV/particle) on coverslips (Fig.~\ref{figS-exsetup}h) and the ND-labeled worms in the antenna-integrated glass bottom dish (Fig.~\ref{figS-exsetup}i). 
The sample temperature ($T_{\rm S}$) was varied via direct heat conduction from the oil-immersion microscope objective whose temperature ($T_{\rm obj}$) was controlled by a PID-feedback controller of the foil heater wrapping the objective (Thorlabs, HT10K \& TC200, temperature precision: $\pm 0.1 \ \si{\degreeCelsius}$).
The immersion oil was Olympus Type-F.
$T_{\rm S}$ was calibrated in the following manner: (1) a tiny flat Pt100 resistance temperature probe (Netsushin, NFR-CF2-0505-30-100S-1-2000PFA-A-4, $5 \times 5 \times 0.2 \ {\rm mm}^{3}$) was tightly attached to the sample coverslip by a thin layer of silicone vacuum grease between the probe and the coverslip. (2) The probe was completely covered by aluminum tape whose edges were glued to the base coverslip. (3) In this thermal configuration, $T_{\rm obj}$ was varied while monitoring $T_{\rm S}$. We obtained the following relation: $T_{\rm S} = 1.84717 + 0.9225 T_{\rm obj}$ (Fig.~\ref{figS-Tscalib}).
The temperature probe was read by a high-precision handheld thermometer (WIKA, CTH7000, precision: 0.01$\si{\degreeCelsius}$). 
During the calibration measurement, the room temperature ($T_{\rm air}$) was monitored using a data logger (T\&D, TR-72wb, precision: 0.5 $\si{\degreeCelsius}$), and we confirmed that $T_{\rm air}$ fluctuates within only $\pm 0.5 \si{\degreeCelsius}$ over 12 h.
Note that $T_{\rm obj}$ was monitored directly on top of the foil heater.

\subsection*{4-point analysis of ODMR signals}
In the 4-point ODMR measurements~\cite{kucsko2013nanometre}, fluorescence intensities at four frequency points ($I_1$ to $I_4$ for $\omega_1$ to $\omega_4$) on CW-ODMR spectra were measured. 
To determine these frequencies, we measured the entire CW-ODMR spectral shape and then applied a fit to the sum of two Lorentzian functions to indicate the ODMR spectrum and  zero-field splitting $D$ (see Fig.~\ref{figS-exsetup}h, i).
Accordingly, the two linear slopes of the ODMR spectrum were recognized via linear fits and the four frequency points, two on each slope, were uniformly distributed, i.e., equally distanced with $\delta \omega$ over the extent of the slopes. 
Here, $\omega_-$ and $\omega_+$ were centered on $D$ such that $I(\omega_-)=I(\omega_+)$. 
$I_1$ to $I_4$ were then given by 
\begin{equation}
\begin{split}
    I_1 & = I(\omega_-)+\gamma_1 \left[-{\delta \omega}+{\delta \beta}+\delta T  \left(\frac{dD}{dT}\right) \right], \\
    I_2 & = I(\omega_-)+\gamma_1 \left[+{\delta \omega}+{\delta \beta}+\delta T \left(\frac{dD}{dT}\right) \right], \\
    I_3 & = I(\omega_+)+\gamma_2 \left[-{\delta \omega}-{\delta \beta}+\delta T \left(\frac{dD}{dT}\right)\right], \\
    I_4 & = I(\omega_+)+\gamma_2 \left[+{\delta \omega}-{\delta \beta}+\delta T \left(\frac{dD}{dT}\right) \right], \\
\end{split}
\label{eq2}
\end{equation} 
where $\gamma_1$ and $\gamma_2$ depict the slopes of the two linear domains. ${\delta \beta}$ is an unknown static magnetic field~\cite{kucsko2013nanometre} but assumed to be zero in the present study.
Note that the splitting of the ODMR dip is due to the interference of the $\rm{^3A}$ spin states and lattice strains~\cite{matsuzaki2016optically,fujiwara2016manipulation,simpson2017non}. 
The splitting does not affect thermometry precision or accuracy in these 4-point measurements.
By assuming that $|\gamma_1|$ and $|\gamma_2|$ are equal (see Supplementary Information for the error of this simplification), the temperature estimate ${\Delta T_{\rm NV}}$ was given by 
\begin{equation}
 \Delta T_{\rm NV} = \delta \omega \left( \frac{dD}{dT} \right)^{-1} \frac{(I_1+I_2)-(I_3+I_4)}{(I_1-I_2)-(I_3-I_4)}.
\label{eq1}
\end{equation}
The total photon counts $I_{\rm tot}$ was obtained by the following equation:
\begin{equation}
    I_{\rm tot} = 1.05 \times (I_1 + I_2 + I_3 + I_4), 
\end{equation}
where the factor of 1.05 was the correction factor accounting for the time interval of 5 $\si{\us}$ during which no photons were detected.

 \subsection*{Error correction of DAQ-board counters}
The systematic errors of the counters in the DAQ system were corrected by measuring the dependence of the counter values ($I_1$ to $I_4$) on the NV fluorescence intensity  (Fig.~\ref{figS-calib}a), where the fluorescence intensity was varied by controlling the laser power.
Two sets of the differences of two counters with the same ODMR depth, $I_1-I_4$ and $I_2-I_3$, were plotted as functions of $I_4$ and $I_3$, respectively and were fitted with second-order polynomials (Fig.~\ref{figS-calib}b).
The counter values of $I_3$ and $I_4$ were corrected using these polynomial functions depending on their values.
With this error-correction, the effect of the systematic error was successfully canceled, as shown in Fig.~\ref{figS-calib}c, 
because intentional step variations in the photon counts ($I_{\rm tot}$) caused artefacts of the estimation without the error-correction, for a constant real temperature.

\subsection*{Particle tracking}
For particle tracking, the piezo stage was scanned in the $xyz$ directions while measuring the ND fluorescence intensity. 
The obtained cross-sections of the point spread function along the $xyz$ axes were fitted with a Gaussian function to determine the $xyz$ positions for re-positioning. 
The piezo stage was moved smoothly to the re-positioning point by five steps of $\sim$ 20 nm every 2 ms.
Re-positioning took 3 s and was performed every 4 s during the 4-point measurement.
Note that temperature estimation was performed every 0.5--1.0 s during the 4 s.
Particle tracking precision was discussed in the Supplementary Information.
Sometimes, ND particles moved beyond the maximal range of particle tracking, particularly in the \textit{in-vivo} experiments shown in Fig.~\ref{fig5} because of the worms' movements. In these cases, the same ND particles were searched and manually moved back to the focus based on the wide-field fluorescence image before the movement.

\subsection*{Determination of precision and accuracy}
The accuracy of $\Delta T_{\rm NV}$ was determined by first adding an offset to $\Delta T_{\rm NV}$ to match $T_{\rm S}$ and by taking the root mean square (RMS) of $T_{\rm S}-T_{\rm NV}$ (Fig.~\ref{figS-precacc}a and Fig.~\ref{figS-invivo-prec}b).
The upper bound of the RMS in the steady state was considered to be the accuracy in this study.
Note that the fluctuation of $T_{\rm air}$ deviated 
$\Delta T_{\rm NV}$ from $T_{\rm S}$, which overestimated the accuracy value.
The precision ($\sigma_{\rm p}$) was determined by taking the standard errors of 20 sampling points of $\Delta T_{\rm NV}$ that were recorded for 38 s (Fig.~\ref{figS-stab} and Fig.~\ref{figS-invivo-prec}a). 
Because this duration comprised 19.4-s integration time ($\delta t_{\rm int}$) and 18.6-s re-positioning time, the sensitivity ($\eta_{\rm T}$) could be calculated as $\eta_{\rm T} = \sigma_{\rm p} \times \sqrt{2 \delta t_{\rm int}}$.
Note that $\Delta T_{\rm NV}$ in the \textit{in-vivo} experiments shown in Figs.~\ref{fig4} and \ref{fig5} has a variation of $\sim \pm$ 8\% propagated from the error of $dD / dT$ (see Supplementary Information).

\subsection*{Highly water-dispersible PG grafted NDs}
ND was grafted with PG according to the method reported previously~\cite{zhao2011chromatographic} (Fig.~\ref{figS-pgnd}a). 
In brief, 5 $\si{mg}$ of hydroxylated NDs with a median diameter of 168 nm (ND-NV-100nm-OH, 900 NV/particle, Ad\'{a}mas Nanotechnologies) was mixed with 5 $\si{mL}$ of glycidol (Alladin chemical, Shanghai), and then bath sonicated at 20 $\si{\degreeCelsius}$ for 30 min. 
The suspension was magnetically stirred at 140 $\si{\degreeCelsius}$ for 20 h.
The resulting yellowish gel was cooled to room temperature and diluted with 20 $\si{mL}$ of water through bath sonication. 
After removing unbound PG by centrifugal filtration (Amicon Ultra-15, MWCO 100K) and washing with water several times, the resulting PG-ND (6.8 $\si{mg}$) was recovered and redispersed in water (1.88 $\si{g} \cdot \si{mL}^{-1}$). 
The PG-ND dispersion showed particle size of 181 nm in the dynamic light scattering (Fig.~\ref{figS-pgnd}b).

\subsection*{ND Labeling \textit{C. elegans} worms}
The wild-type \textit{C. elegans} strain Bristol N2 was obtained from the Caenorhabditis Genetics Center (CGC, Minneapolis, MN). 
\textit{C. elegans} was  maintained using standard techniques~\cite{brenner1974genetics}. Young adult hermaphrodites were
injected by using the standard procedures~\cite{mello1991efficient}, with some modifications.
In brief, glass capillaries (ND-1, Narishige) were pulled using a pipette puller (Olympus, PC-100). 
Needles were filled with the PG-ND dispersion. 
They were mounted on a manipulator (Narishige, MN-4) and pressurized through an injection system (Narishige, IM-31). 
Worms were immobilized on the agarose injection pads that were applied with paraffin oil (Wako, Japan). 
The microscope used for the injection was equipped with differential interference contrast (Olympus, IX73). 
The PG-ND dispersion was injected into the distal arm of the gonad (Fig.~\ref{figS-celegans-parts}).
The injected worms recovered on the bacteria-seeded NGM plates for more than a day.

\subsection*{Transfer of worms and adding chemical stimuli}
The ND-labeled worms were transferred from the culture dishes onto the agar pads prepared on small pieces of glass substrates (Fig.~\ref{figS-exsetup}j, k).
A small aliquot of sodium azide solutions (50 mM) was dropped for anesthesia.
The agar pads that held the worms were placed on the antenna-integrated glass-bottom dishes, while the worm position was adjusted near the microwave antenna. 
10 mM FCCP (carbonyl cyanide-\textit{p}-trifluoromethoxyphenylhydrazone) (Sigma-Aldrich, Japan)  in dimethyl sulfoxide (DMSO) (Wako, Japan) was diluted to 0.6\% (v/v) with M9 buffer (5 mM potassium phosphate, 1 mM CaCl$_2$, 1 mM MgSO$_4$) to give 60-$\mu$M FCCP solution. 
The 0.6\% (v/v) DMSO/M9 buffer solution without FCCP was used for the control experiment. 
 Small droplets of FCCP solution and vehicle-control buffer solution were placed near the agar pads so as to spontaneously spread into the sandwiched region.
These measurements were performed at 23 $\si{\degreeCelsius}$ with a periodic fluctuation of $\pm 0.5 \si{\degreeCelsius}$ every 40 min.
For the particle tracking measurement shown in Fig.~\ref{figS-tracking}, a commercial confocal microscope (Leica, Germany) was used with a setup of laser excitation at 552 nm and emission filter of 645--700-nm transmission window.

\subsection*{Statistical analysis of time profile of $\Delta T_{\rm NV}$}
To quantify the response curves of the increase in temperature, we chose peak height, latency, and duration of the response curves as the characteristic parameters. 
Because $\Delta T_{\rm NV}$ sometimes exhibits long-term baseline drifts during the measurements and sudden jumps at the time of the droplet addition (either FCCP or control), these artefacts were linearly subtracted. c
These compensated data were low-pass-filtered by the Lowess filter with the span of 0.1 to extract the envelope of the response curve.
The above three parameters were determined from these curves.
Note that the baseline drifts and jumps of $\Delta T_{\rm NV}$ only occur in worms and not for NDs on coverslips in air.
To obtain further details on these phenomena, we measured the CW-ODMR and fluorescence spectra of NDs before and after the long-term drifts and sudden jumps as shown in Fig.~\ref{figS-driftjump}.
The CW-ODMR spectra were shifted in agreement with the 4-point measurement data. However, slight changes in spectral shapes were also observed (see Supplementary Information for the interpretation of the statistical data).

\subsection*{Data availability}
The datasets analyzed during the current study are available from the corresponding authors on reasonable request.

\clearpage
\section*{Acknowledgment}
We thank O. Shenderova (Ad\'{a}mas Nanotechnologies Inc.) for kindly providing ND-NV-100nm-OH, and J. Choi, M. D. Lukin, T. Matsubara, P. Maurer, K. Xia, K. Yoshizato, and H. Zhou for fruitful discussions.
This work is supported in part by Osaka City University Strategic Research Grant 2017 \& 2018 (M.F., E.K.N., Y.S., A.D.), Murata Science Foundation (M.F., E.K.N.), and JSPS-KAKENHI (M.F.: 16K13646, and 17H02741, Y.S.: 19K14636, N.K.:17H02738).
M.F. acknowledges funding by MEXT-LEADER program and Sumitomo Research Foundation.
O.B., N.S., A.D. acknowledge funding by the Deutsche Forschungsgemeinschaft DFG (FOR 1493).

\section*{Author Contributions}
M.F., E.K.N. conceived the idea. 
M.F., E.K.N., Y.S. designed the research project.
A.D., M.F., Y.N., K.O., Y.U., YS.T., N.S., O.B. developed the real-time thermometry system and accessories.
S.S., YK.T., E.K.N. performed worm preparation and their ND labeling.
M.F., S.S., K.S. performed the \textit{in-vivo} thermometry experiments.
Y.S., M.F. analyzed the temperature estimation protocol.
Z.L., N.K. prepared the PG-NDs. 
All authors participated in the discussion and writing of the paper.

\section*{Notes}
The authors declare no competing interest. 

\clearpage

% \onecolumngrid
% \appendix
\clearpage
\onecolumngrid
\section*{Supplementary Information}
\renewcommand{\thefigure}{S\arabic{figure}}
\setcounter{figure}{0}    

\section*{Effect of assuming that the two linear slopes of the ODMR spectrum are equal}

In the Methods section, to obtain 
Eq. 1 
from Eq. 2, 
$|\gamma_1|$ and $|\gamma_2|$ were assumed to be equal. 
However, they are slightly different ($\sim 5.0 \%)$, as summarized in Table~\ref{tbl:table1}.
This difference may affect temperature estimates, particularly in the \textit{in-vivo} experiments, because slight changes of the ODMR spectra were observed in worms 
(Fig.~S10). 

\begin{table*}[th!]
\small
  \caption{Variation of $\gamma_1$ and $\gamma_2$ in the 4-point selection process.}
  \label{tbl:table1}
 \begin{tabular}{cccc}
   \hline
    Sample ND & $\gamma_1$ [MHz$^{-1}$] & $\gamma_2$ [MHz$^{-1}$] & Difference [\%] \\
    \hline
    \textbf{1}  &   -4.821    &   4.594     &   4.7 \\ 
    \textbf{2}  &   -8.807    &   9.112     &   3.3 \\ 
    \textbf{3}  &   -6.152    &   5.825     &   5.6 \\ 
    \textbf{4}  &   -9.326    &   8.825     &   5.4 \\ 
    \textbf{5}  &   -4.440    &   4.194     &   5.9 \\ 
    \hline
 \end{tabular}
\end{table*}

\section*{Temperature dependency of the zero-field splitting and fluorescence intensity of NV centers in NDs and its error propagation to the temperature estimation}
In the main text, we determined the  temperature dependency of $D$ by measuring five NDs, which was $dD/dT  = - 65.4 \pm 5.5 \ \si{\kHz} \cdot \si{\degreeCelsius}^{-1}$. 
This value is relatively smaller than the previously reported values of approximately $- 74 \ \si{\kHz} \cdot \si{\degreeCelsius}^{-1}$~\cite{acosta2010temperature,toyli2012measurement,chen2011temperature}. 
Although the exact reason for this discrepancy is not known, we should note two possibilities.  
(1) Material inhomogeneity: The above reported values were determined for the NV centers in bulk diamonds and have a significant variation from $-71$ to $-84\  \si{\kHz} \cdot \si{\degreeCelsius}^{-1}$ depending on the sample~\cite{acosta2010temperature,toyli2012measurement,chen2011temperature}. 
It is known that the NV centers in NDs show greater variation of the zero-field splitting than the bulk because of various factors including crystal strains and surface states~\cite{foy2019wide,plakhotnik2014all}.
Furthermore, the temperature dependence of zero-field splitting for NDs has not been thoroughly studied for different types of ND samples. 
(2) Calibration difficulty: In the present study, we calibrated the sample temperature ($T_{\rm S}$) by using the tiny flat Pt100 temperature probe tightly attached to the coverslip with heat conducting tape. 
While this method is usually used for measuring the surface temperature, $T_{\rm S}$ might not be the same as the exact temperature that NDs are feeling on the coverslip surface.
Because the NDs are surrounded by air in contrast to the Pt100 probe covered by the heat conducting tape, NDs can feel temperatures lower than $T_{\rm S}$. In this case, $dD/dT$ can be underestimated.

The temperature dependence of the fluorescence intensity, $I_{\rm tot}^{-1} \ dI_{\rm tot}/dT = -3.9 \pm 0.7 \ \% \cdot \si{\degreeCelsius}^{-1}$, 
is also larger than the previously reported values for NDs; the previous study reports the same variation in the fluorescence intensity 
for the change of $\sim 150 \si{\degreeCelsius}$ (with a large variation of 70--270  $\si{\degreeCelsius}$). 
Besides the temperature dependency of the NV centers, the overall optical throughput of the microscopy system may be related to the observed fluorescence variation. 
The imaging property of the microscope objective is optimized for room temperature at approximately 20 $\si{\degreeCelsius}$.
The temperature variation may change the diffraction pattern and the back focal pattern of the laser and fluorescence, which finally affects the optical throughput at the pinhole because of spatial filtering. 
It should be noted that the fluorescence intensity is strongly affected by the optical transmission inside the worm's body and cannot be used independently as a temperature indicator.
The clear observation of the thermal events in $\Delta T_{\rm NV}$ in live worms is by itself a strong indication of the temperature change in the worms.

The standard deviation of $dD/dT$ is propagated to the temperature estimate $\Delta T_{\rm NV}$.
By denoting error $\delta D'$, the error propagation to $\Delta T_{\rm NV}$ can be calculated as follows: 
\begin{equation}
    \delta \Delta T_{\rm NV} = \delta D' \left(\frac{dD}{dT} \right)^{-1}  \Delta T_{\rm NV}, 
    \nonumber
\end{equation}
which results in $\sim$ 8 \% of the measured $\Delta T_{\rm NV}$.

\section*{Particle tracking precision}
Particle tracking precision in this study was mainly limited by the step size of the re-positioning process.
For re-positioning, the piezo stage was scanned in the $xyz$ directions with a step size of 32 nm to obtain the point spread function. 
The one-dimensional cross-section was fitted with a Gaussian function to determine the central peak position. 
In general, well-isolated single peaks allow for accurate positioning of $\sim$ 30 nm 
(Fig.~4), 
whereas distorted or double peak functions deteriorate positioning accuracy, particularly in live worms 
(Fig.~8).
Furthermore, in live worms, other NDs moving around sometimes get closer to the focus and generate a second peak in the point spread function of the locked ND, which in many cases prohibits particle tracking.
Note that particle tracking is a feedback process to maximize the fluorescence counts of NDs that are likely to move away from the focus and can be coupled with variations of fluorescence intensity derived from temperature changes in NV centers depending on feedback protocols.
While the present feedback protocol is not significantly coupled with the temperature estimation because of the constant repositioning time, 
it is somewhat coupled with the thermometry, which becomes significant in the low photon regime of $I_{\rm tot} < 0.5$ Mcps.
We therefore adjusted the photon counts to always be larger than this threshold (ideally $I_{\rm tot} > 0.75$ Mcps, as in 
Fig.~3 in the main text) 
to exclude any measurement artefacts. 

\section*{Interpretation of the statistical analysis of \textit{in-vivo} thermogenic responses}
Although we clearly observed a temperature increase inside worms via the time profiles of $\Delta T_{\rm NV}$ (Fig.~4) and their correlation plot with the integrated area of the response curves (Fig.~S10), its quantitative interpretation is challenging at present because of incomplete understanding of the sensory behaviors of NV centers in live worms and lack of complete information regarding \textit{C. elegans} thermogenesis. 
First, the observed drifts and jumps in $\Delta T_{\rm NV}$ (also confirmed as shifts in CW-ODMR spectra, as in Fig.~S10), 
cannot be explained well at present because spin properties of NDs in living worms have not been well studied.
Second, the mean values of $\Delta T_{\rm NV}$ for the FCCP stimulation and control experiments cannot be directly compared, for example, by subtracting the two cases, because the time profiles of $\Delta T_{\rm NV}$ were very different. 
For example, the present analysis shows +0.7$\si{\degreeCelsius}$ at 23 min for the vehicle control in 
Fig.~4d in the main text, 
although its interpretation is not straightforward.
Third, the possible variation of $dD/dT$ of NDs can affect the temperature values. 
While this study uses $dD/dT  = - 65.4 \ \si{\kHz} \cdot \si{\degreeCelsius}^{-1}$, as mentioned above, there is a range of $dD/dT$ from -50 to -100 $\si{\kHz} \cdot \si{\degreeCelsius}^{-1}$ reported~\cite{foy2019wide}.
Without determining $dD/dT$ for each measured ND \textit{in situ}, the quantitative interpretation of the temperature response curve is challenging.
Fourth, it is still not clear how physiological thermogenesis, including shivering and non-shivering thermogenesis, occurs in live worms. 
The slight temperature increase in the control experiments may be ascribed to shivering thermogenesis if the temperature estimate of such small values is confirmed to be real; yet one is not able to conclude the shivering thermogenesis only by this temperature measurement data, requiring more comprehensive experiments.  

In the same way, the present data cannot be directly compared with those from a previous report on micro-machined calorimeter measurements, where metabolic heat generation by a few hundred \textit{C. elegans} worms by FCCP stimulation with a relatively low concentration was measured~\cite{krenger2018dynamic}, because of the differences in experimental parameters, such as FCCP concentration, treatment time, and population inhomogeneity of the temperature responses.
More thorough analyses of the NV sensory behaviors in live worms and the observed temperature increases are necessary to understand the correspondence of the present data with that of the previous report. 

\clearpage
\section*{Supplementary Figures}
\begin{figure}[h!]
 \centering
 \includegraphics[scale=0.8]{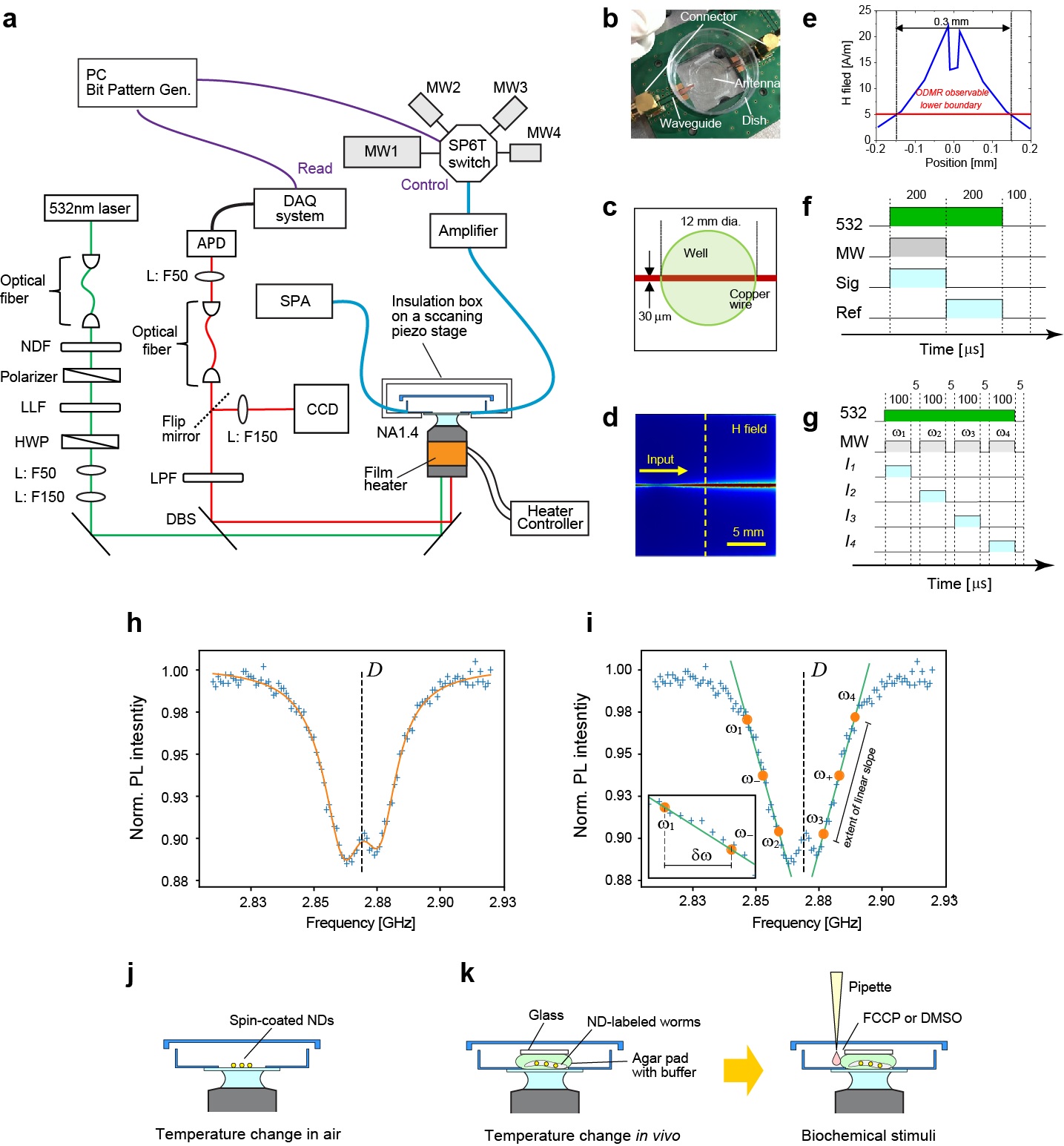}
  \caption{Experimental setup. (a) Schematic of the experimental setup for the optical layout and microwave circuit. NDF: neutral density filter. LLF: laser line filter. HWP: half-wave plate. L: lens. DBS: dichroic beam splitter. LPF: long pass filter. CCD: charge-coupled device camera. APD: avalanche photodiode. SPA: spectrum analyzer. MW: microwave source. DAQ: data acquisition board. (b) Photograph of antenna-integrated glass bottom dish on a printed circuit board holding a coplanar waveguide. (c) Geometry of electromagnetic simulation of microwave linear antenna based on a finite element method. (d) Magnetic field intensity map. (e) 1D cross-sectional plot of the magnetic field along the yellow-dashed line. Lower boundary of 5 $\si{A}\cdot \si{m}^{-1}$ was determined experimentally. (f) Pulse control sequences for CW-ODMR measurements and (g) four-point measurements. 532: green laser. MW: microwave. Sig: signal for $I_{\rm PL}^{\rm ON}$. Ref: reference for $I_{\rm PL}^{\rm OFF}$. $\omega_1$ to $\omega_4$ are the four frequencies used for the four-point measurements. 
 (h) ODMR spectra fitted using the sum of two Lorentzians centered at the zero-field splitting frequency $D$ and (i) two linear functions to the slopes. Selected frequencies of $\omega_1$ to $\omega_4$ and their intermediates, $\omega_{-}$ and $\omega_{+}$), are indicated by orange spheres and are separated on each slope by $\delta \omega$, as shown in the inset. 
  (j) Schematic of NDs on coverslip in air and (k) of ND-labeled worms.
 }
 \label{figS-exsetup}
\end{figure}

\clearpage

\begin{figure}[h!]
 \centering
 \includegraphics[scale=0.85]{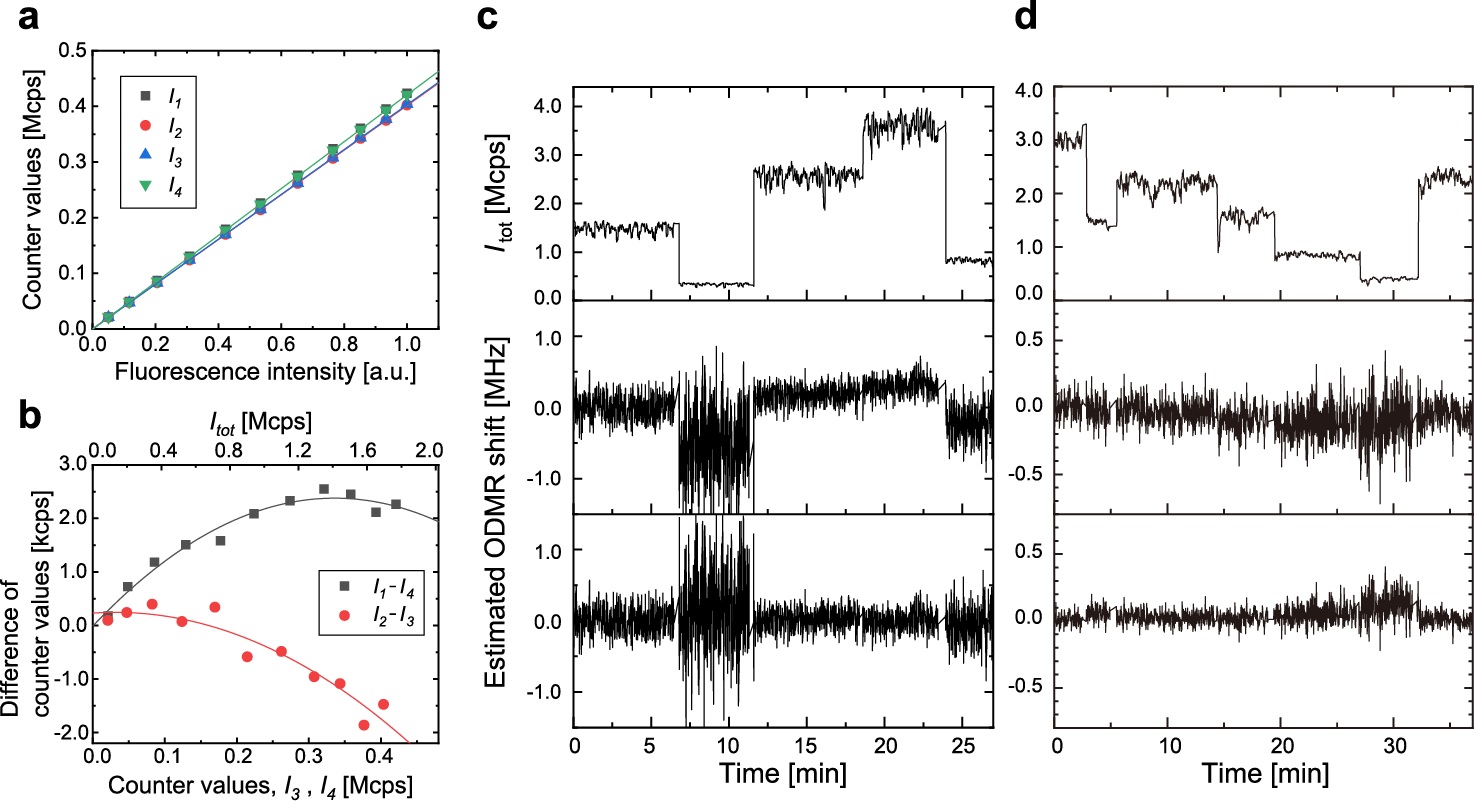}
 \caption{Variations of photo-responsivity of counters and the error-correction results. (a) Photon counts of each counter from $I_1$ to $I_4$ as a function of ND fluorescence with their linear fits. $I_1$, $I_4$ and $I_2$, $I_3$ show similar increases reflecting the difference in the ODMR contrast. However, a very small difference was observed between $I_1$ and $I_4$ ($I_2$ and $I_3$) as an instrumental artefact.
 (b) Difference in the counter values between the two sets of counters, namely  $I_1 - I_4$ and $I_2 - I_3$, as functions of $I_4$ and $I_3$, respectively. 
 Solid lines represent second-order polynomial fits to the data.
 (c) Time profiles of photon counts of all the counters ($I_{\rm tot}$, top) over 27 min with intentional variations in NV fluorescence intensity by the laser intensity control (top). 
 The corresponding time profiles of the estimated ODMR shift without (middle) and with (bottom) the error correction. The sampling time is 500 ms. $T_{\rm S}$ is set at 36 $\si{\degreeCelsius}$. (d) Time profiles of photon counts of all the counters ($I_{\rm tot}$, top) over 27 min with intentional variations in NV fluorescence intensity by varying the fluorescence intensity control (top). 
 The corresponding time profiles of the estimated ODMR shift without (middle) and with (bottom) the error correction. The sampling time is 500 ms. $T_{\rm S}$ is set at 36 $\si{\degreeCelsius}$.
 }
 \label{figS-calib}
\end{figure}

\begin{figure}[h!]
 \centering
 \includegraphics[scale=1.0]{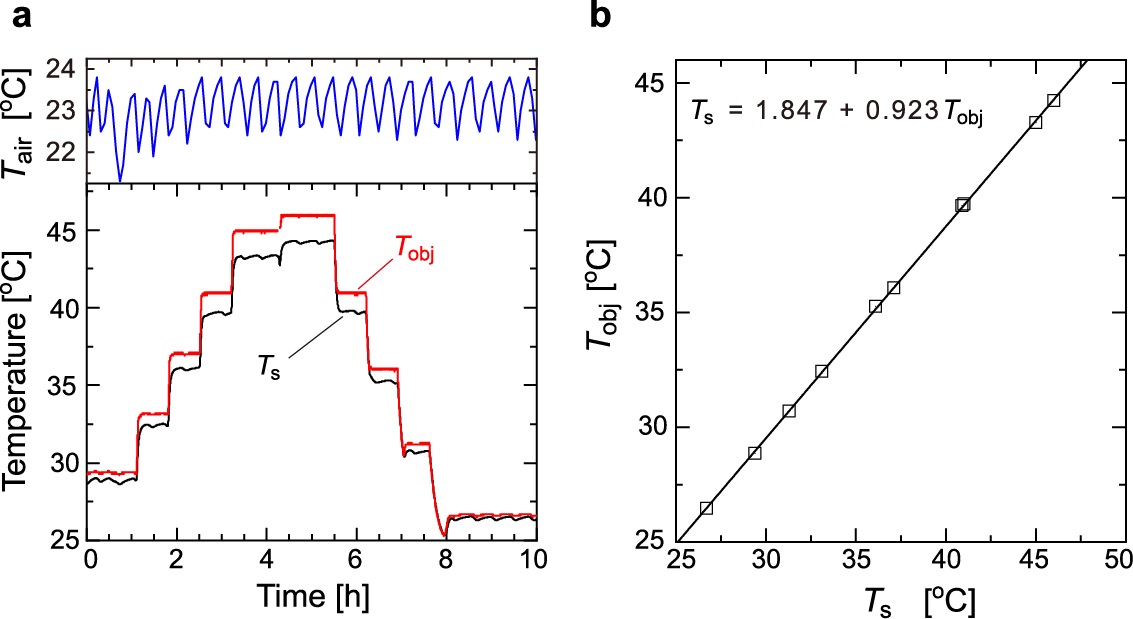}
 \caption{
 Calibration of the sample temperature ($T_{\rm S}$) with that of the microscope objective ($T_{\rm obj}$). (a) Temperature profiles of room temperature ($T_{\rm air}$, blue), $T_{\rm obj}$ (red), and $T_{\rm S}$ over 10 h. (b) The obtained relation between $T_{\rm obj}$  and $T_{\rm S}$ with the linear fit. $T_{\rm S} = 1.847 + 0.923 T_{\rm obj}$ is obtained.
 }
 \label{figS-Tscalib}
\end{figure}

\begin{figure}[h!]
 \centering
 \includegraphics[scale=0.9]{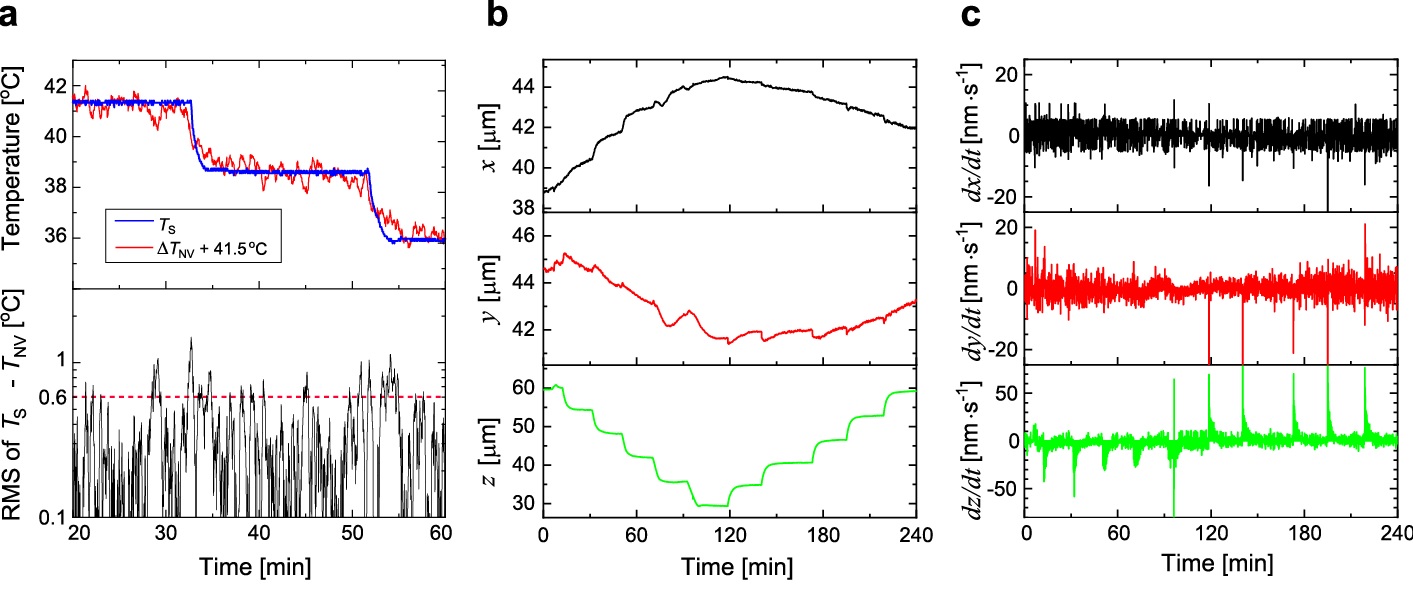}
 \caption{Accuracy of $\Delta T_{\rm NV}$ and tracking profile for the ND on the coverslip in air. (a) Time profiles of moving-averaged $\Delta T_{\rm NV}$ with $T_{\rm S}$ (top) and of the RMS of $ T_{\rm S}-T_{\rm NV}$ (bottom).
 The sampling rate is 1 s. The red dashed line indicates the accuracy threshold.
 (b) Time profiles of ND positional tracking in the $xyz$ axes during stepwise temperature changes over 240 min.
 (c) Time profiles of the speed of the ND positional move (first derivative of the positional plot).
 }
 \label{figS-precacc}
\end{figure}

\begin{figure}[h!]
 \centering
 \includegraphics[scale=0.95]{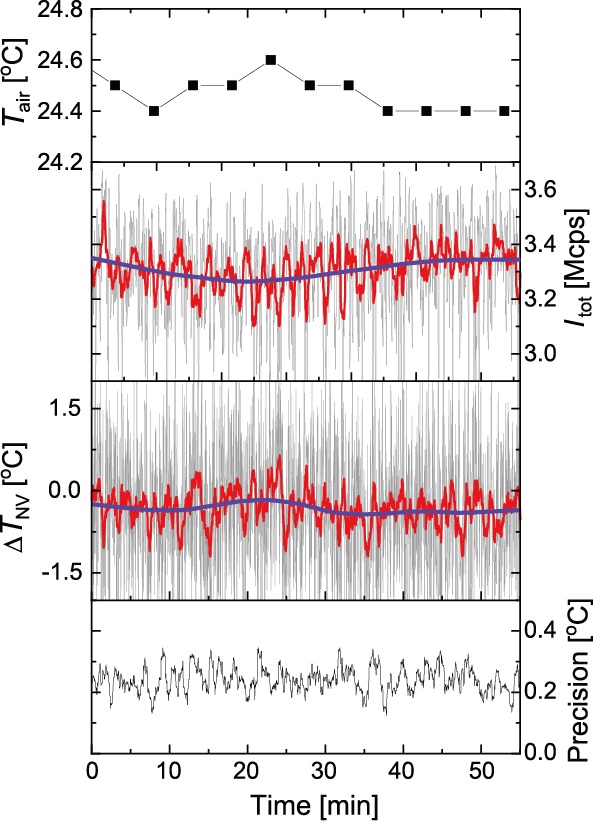}
 \caption{Stability of $\Delta T_{\rm NV}$ for NDs on the coverslip in air. Time profiles of $T_{\rm air}$ (top), $I_{\rm tot}$ (2nd top), $\Delta T_{\rm NV}$ (2nd bottom), and 19.4-s-integrated precision (20 sampling points) (bottom).
 Gray: 1-s sampling data. Red: moving averaged data of 20 sampling points. Violet: Lowess filter smoothed data to extract long-term change. $T_{\rm S}$ is set at 36 $\si{\degreeCelsius}$.
 }
 \label{figS-stab}
\end{figure}

\begin{figure}[p!]
 \centering
 \includegraphics[scale=1]{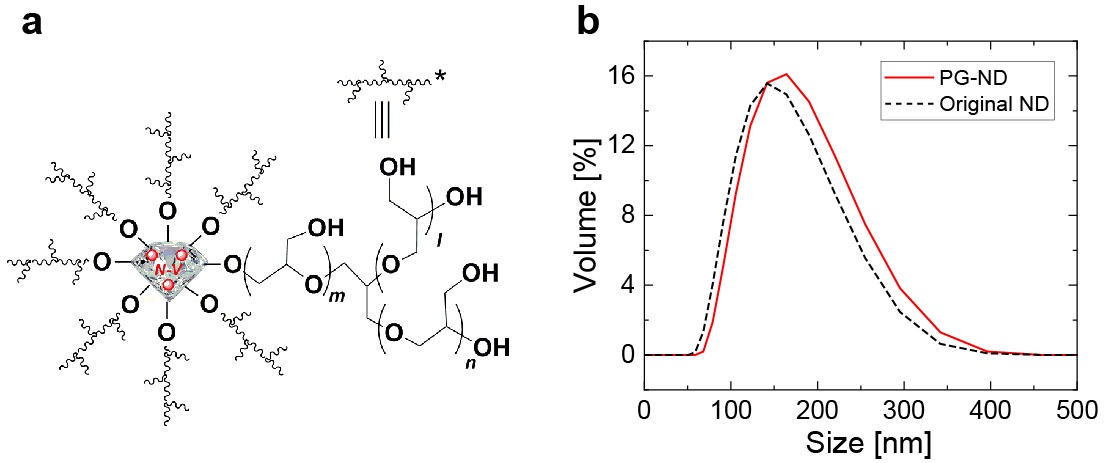}
 \caption{(a) Schematic of PG-NDs and (b) its dynamic light scattering data with that of original NDs before the surface functionalization.}
 \label{figS-pgnd}
\end{figure}

\begin{figure}[h!]
 \centering
 \includegraphics[scale=0.9]{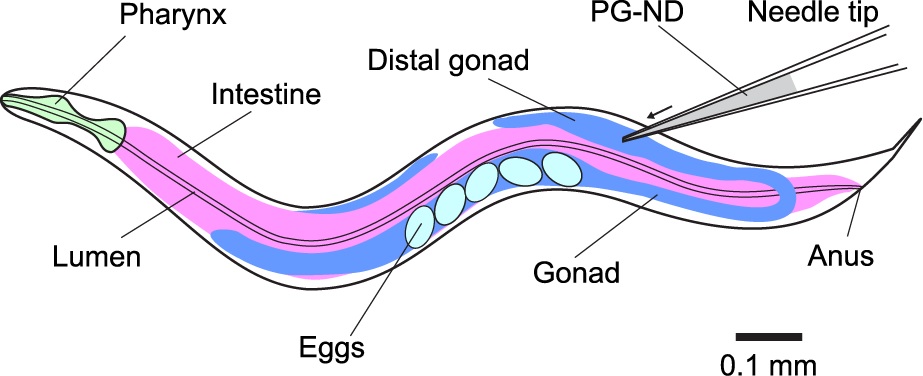}
 \caption{Anatomy of an adult \textit{C. elegans} for microinjection of PG-NDs into the gonad. Anatomical structure of an adult hermaphrodite, left lateral side. The scale bar indicates 0.1 mm. PG-ND suspension was microinjected into the distal arm of the gonad. PG-NDs were dispersed in the distal gonad and oocytes. 
}
 \label{figS-celegans-parts}
\end{figure}

\begin{figure}[h!]
 \centering
 \includegraphics[scale=1]{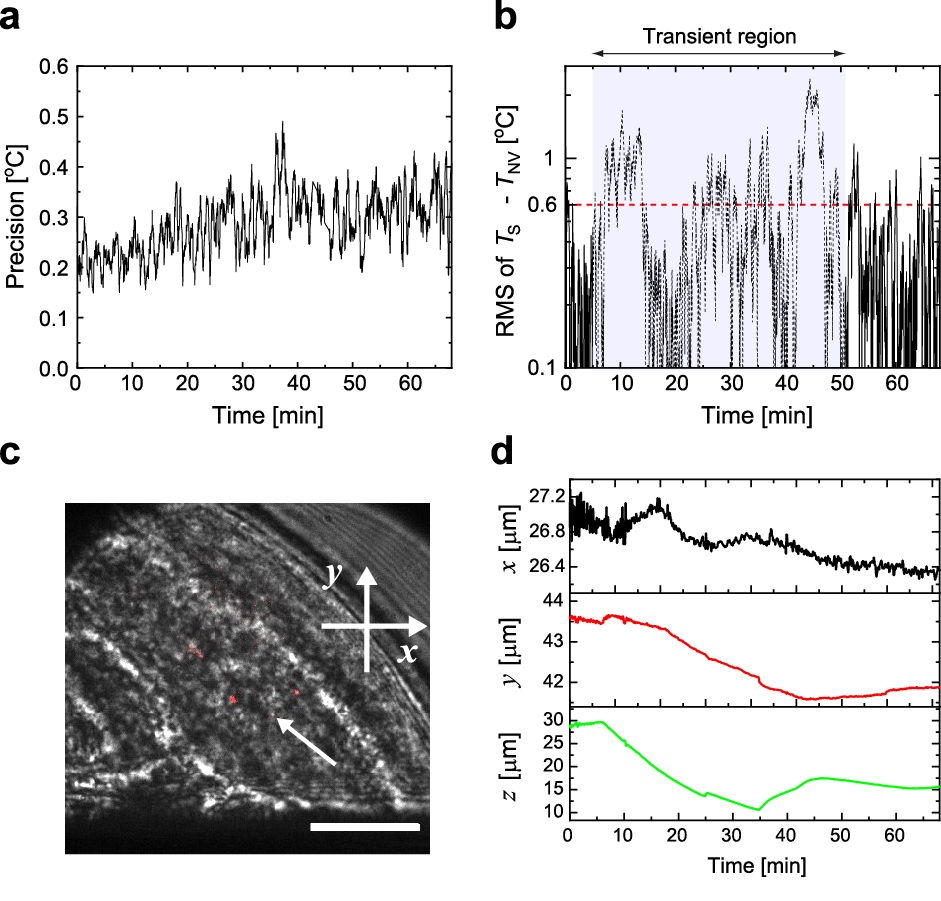}
 \caption{Precision, accuracy, and positional data of \textit{in-vivo} thermometry. Time profiles of (a) precision (20 sampling points) and (b) of the RMS of $T_{\rm S}-T_{\rm NV}$. 
 (c, d) Photograph indicating the $xyz$ axes and time profile of the positional tracking in the $xyz$ axes during the measurement. The $z$-axis is perpendicular to the paper. Scale bar: 20 $\si{\um}$.}
 \label{figS-invivo-prec}
\end{figure}

\begin{figure*}[h!]
 \centering
 \includegraphics[scale=1.0]{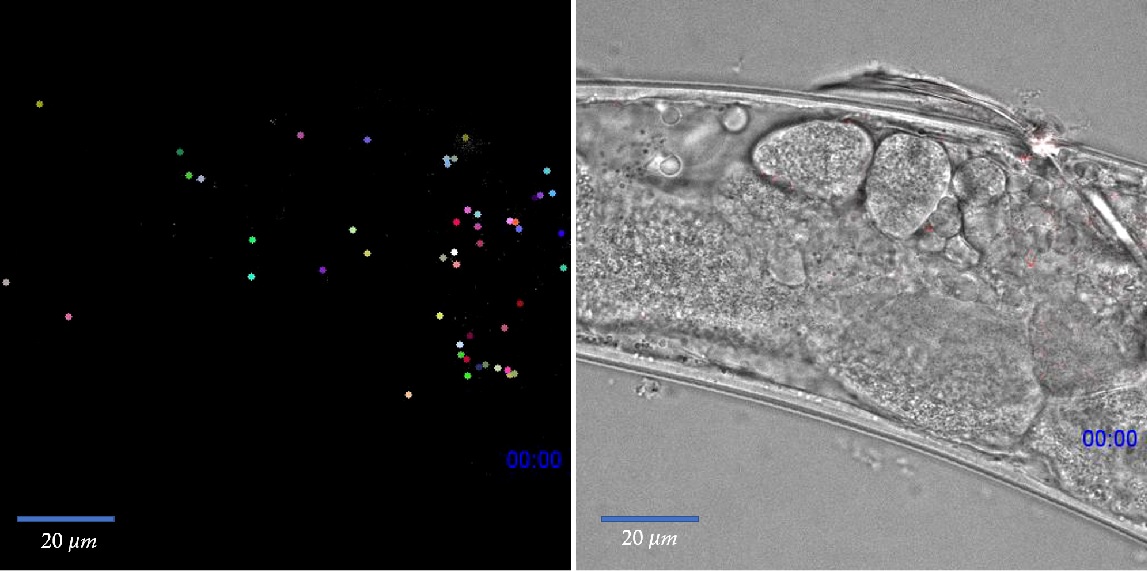}
 \caption{Time-lapse movies of NDs in worms during FCCP stimuli over 50 min.
 Left: trajectories of NDs. Right: Merged image of red and transmission. The montage video is provided in the following link \href{https://www.dropbox.com/s/8bmc0e10e6toyhl/Particle%20Track%20Movie-2.mp4?dl=0
 }{\textcolor{blue}{[MP4 video].}}
 }
 \label{figS-tracking}
\end{figure*}

\begin{figure*}[h!]
 \centering
 \includegraphics[scale=0.9]{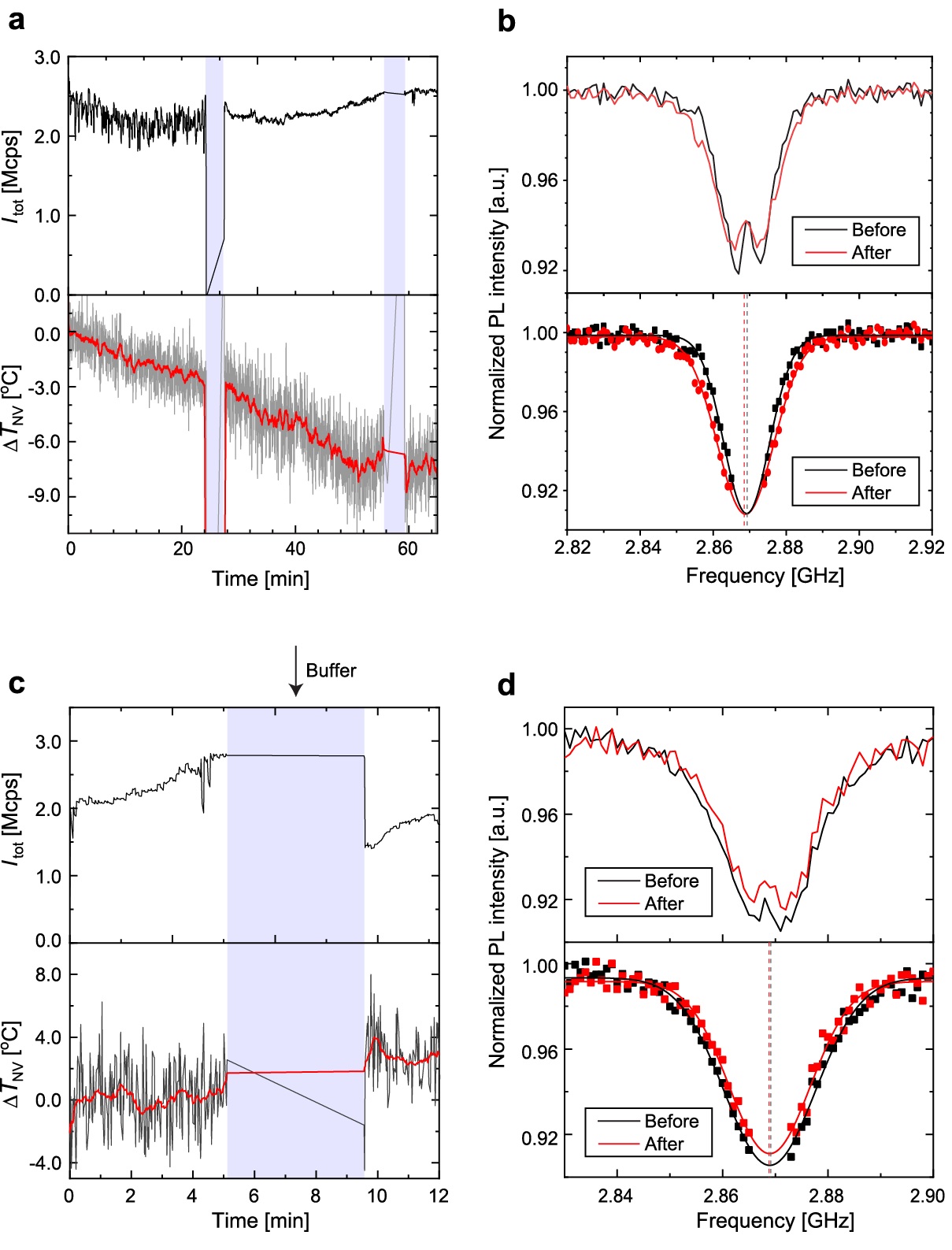}
 \caption{Detailed analysis of the baseline drifts and jumps of $\Delta T_{\rm NV}$ inside worms. (a) Time profile of $\Delta T_{\rm NV}$ over 70 min. The blue shaded regions are periods during which the temperature measurements were not performed. (b) CW-ODMR spectra before and after the baseline drift with Gaussian fitting. The peak splitting structures are omitted to enable curve-fitting as described in Ref.~\citenum{simpson2017non}. (c) Time profile of $\Delta T_{\rm NV}$ when the baseline jump occurs after the addition of buffer. The baseline jumps only occurs after the droplet addition. (b) CW-ODMR spectra before and after the baseline jumps with the Gaussian fitting.}
 \label{figS-driftjump}
\end{figure*}

\begin{figure*}[h!]
 \centering
 \includegraphics[scale=1.0]{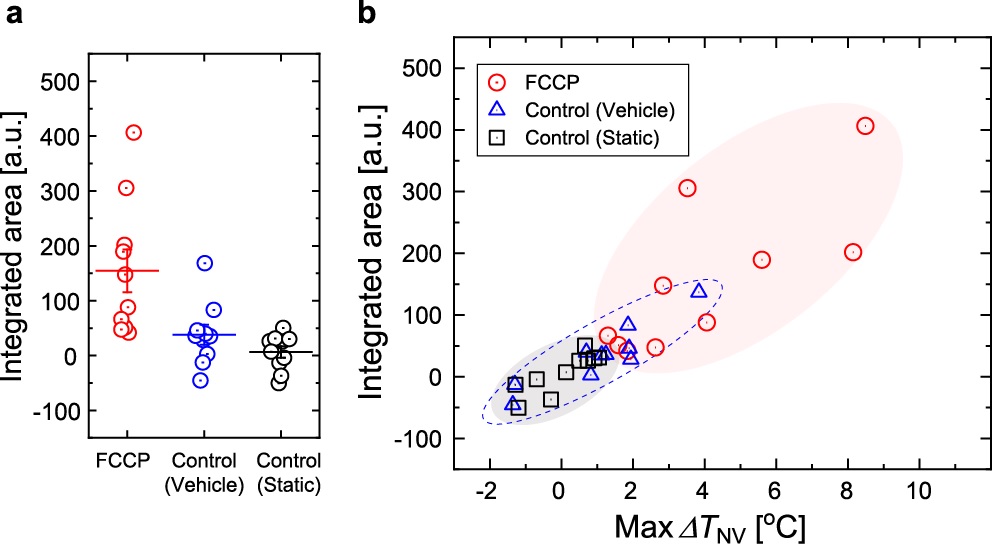}
 \caption{Additional statistical analysis of the temperature response curves. (a) Statistical plots of the integrated area of the response curves for FCCP stimulation, vehicle-control, and static-control (no solution added). $n$ = 10 for all the data.
 The mean values of the integrated area with standard error are ($155 \pm 39$), ($38 \pm 18$), and ($7 \pm 10$) for the FCCP, vehicle-control, and static-control experiments, respectively.
 (b) Correlation plots between the maximum $\Delta T_{\rm NV}$ and integrated area of the temperature response curves for the three cases, FCCP addition (red circles), vehicle control (addition of vehicle without FCCP; blue triangles), and static control (no solution added; black boxes). Clearly, there are three clusters representing the above three types of experiments. The wide distribution of FCCP is prominent probably because of insufficient controllability of the FCCP concentration or dehydration-induced viability.
 }
 \label{figS-correlation}
\end{figure*}

\clearpage
%\bibliography{achemso-demo}
%\bibliographystyle{apsrev4-1}


\end{document}